\documentclass[journal]{IEEEtran}

\usepackage{cite}

\ifCLASSINFOpdf
   \usepackage[pdftex]{graphicx}
\else
   \usepackage[dvips]{graphicx}
\fi
\usepackage{booktabs}
\usepackage{enumitem}
\usepackage{soul} 
\usepackage{xcolor} 
\usepackage{placeins} 
  
\usepackage[cmex10]{amsmath}
\usepackage{siunitx}
\DeclareSIUnit \voltampere { VA } 
\DeclareSIUnit \pu { pu } 

\newcommand{\subm}[2]{\text{#1}_{\text{#2}}}

\usepackage{array}

\ifCLASSOPTIONcompsoc
 \usepackage[caption=false,font=normalsize,labelfont=sf,textfont=sf]{subfig}
\else
 \usepackage[caption=false,font=footnotesize]{subfig}
\fi

\usepackage{float}
\begin{document}
%
\title{Power Oscillation Damping Controllers for Grid-Forming Power Converters in Modern Power Systems}

\author{Elia~Mateu-Barriendos, Onur Alican, 
        Javier Renedo,~\IEEEmembership{Senior Member,~IEEE,}
        Carlos Collados-Rodriguez,  
        Macarena Martin, Edgar Nuño,
        Eduardo~Prieto-Araujo,~\IEEEmembership{Senior Member,~IEEE,}
        and~Oriol~Gomis-Bellmunt,~\IEEEmembership{Fellow,~IEEE,}
\thanks{Elia~Mateu-Barriendos, Onur Alican, Carlos Collados-Rodriguez, Eduardo~Prieto-Araujo and Oriol~Gomis-Bellmunt are with CITCEA-UPC. Javier Renedo, Macarena Martin and Edgar Nuño are with Red Eléctrica - Redeia. Contact: elia.mateu@upc.edu. This is an unabridged draft of paper TPWRS-01501-2024 submitted on 6th September 2024 to
IEEE Transactions on Power Systems and currently under
review.}}


\maketitle


\begin{abstract}
Inter-area oscillations have been extensively studied in conventional power systems dominated by synchronous machines, as well as methods to mitigate them.  Several publications have addressed Power Oscillation Damping (POD) controllers in grid-following voltage source converters (GFOL). However, the performance of  POD controllers for Grid-Forming voltage source converters (GFOR) in modern power systems with increased penetration of power electronics requires further investigation. This paper investigates the performance of GFORs and supplementary POD controllers in the damping of electromechanical oscillations in modern power systems.  This paper proposes POD controllers in GFORs by supplementary modulation of active- and reactive-power injections of the converter and both simultaneously (POD-P, POD-Q and POD-PQ, respectively). The proposed POD controllers use the frequency imposed by the GFOR as the input signal, which  has a simple implementation and it eliminates the need for additional measurements. Eigenvalue-sensitivity methods using a synthetic test system are applied to the design of POD controllers in GFORs, which is useful when limited information of the power system is available. This paper demonstrates the effectiveness of POD controllers in GFOR converters to damp electromechanical oscillations, by small-signal stability analysis and non-linear time-domain simulations in a small test system and in a large-scale power system. 
\end{abstract}



\vspace{-1.0em}
\section{Introduction}
\noindent Stability issues related to  electromechanical oscillations (also known as power oscillations or low-frequency oscillations) have been repeatedly observed in large power systems. In 2016, an undamped oscillation at \qty{0.15}{\hertz} in the Continental European power system was reported by ENTSO-E \cite{entsoe-report-16}. Another oscillatory event at \qty{0.29}{\hertz} was detected in 2017 \cite{entsoe-report-17}. Similarly, a low-damped inter-area oscillation around \qty{0.27}{\hertz} was observed across the Eastern Interconnection of the North American grid also in 2016 \cite{ie2019}. 


 Electromechanical oscillations are typically caused by interactions between Synchronous Generators (SGs) and can be classified as inter-area or intra-area \cite{1994power}. Inter-area oscillations (\qtyrange[range-phrase=--]{0.1}{0.7}{\hertz}) involve groups of generators in different areas oscillating against each other, whereas intra-area oscillations (\qtyrange[range-phrase=--]{0.7}{2}{\hertz}) are associated with local interactions between a few closely-located generators. Traditionally, control strategies like Power System Stabilizers (PSS)  in synchronous machines have proved to be effective in damping such oscillations in power systems dominated by synchronous  generation \cite{kundur-pss,larsen}. 

However, modern power systems are evolving towards an increased penetration of renewable generation, which results in the progressive replacement of SGs by converter-interfaced generation  (CIG), mainly using Voltage Source Converters (VSC). Currently, most extended control of VSC power converters is Grid-Following (GFOL). Nevertheless, Grid-Forming (GFOR) technology is required to deal with power systems with large amounts of CIG~\cite{Carmen_Cardozo_GFM_2024}, even for scenarios with 100 \% of CIG. This transition brings new stability challenges, as the oscillatory behavior of the system becomes also affected by the control structures of the power converters~\cite{Nikos_2021}. In particular, electromechanical modes might be affected by these converter-interfaced units, although their specific impact depends on many factors, such as their location, level of penetration, and mode of operation~\cite{relevance_inertia,inertia_distribution, wind_interarea}. Hence, in recent years other damping strategies based on Power Oscillation Damping (POD) controllers have been proposed for FACTS \cite{comparisonPSS,placement_facts}, wind turbines \cite{pod_review,wind_pod,inout_wpp_pod}, PV generators \cite{lowfreq_pv,pod_pv},  energy storage systems (ESS) \cite{ISanz_2022} and VSC-HVDC systems \cite{hvdc_interarea_survey, hvdc_interarea_control, renedoDesignPOD}. 

However, all the publications above deal with POD controllers in GFOL VSCs. GFOR technology is less mature, and POD controllers in GFOR power converters have received less attention. The work in  \cite{droop_vs_vsm, en15124273, Johan_2023} has explored the inherent damping capability of GFOR converters using power-frequency (P-f) droop and virtual synchronous machine (VSM) schemes~\cite{gfol_gfor, droop_eq_vsm},  highlighting the benefits of the inherent damping capability of GFOR converters, if properly designed. The work in \cite{Mandrile2023} analysed the inherent damping of a GFOR connected to an infinite grid using different control variants and it proposed a lead/lag filter in the VSM scheme to improve its dynamic behaviour. However, the impact on electromechanical oscillations in power systems was not analysed. The work in \cite{tdm_pod} proposed the Transient Damping Method (TDM) in GFOR converters, by adding a transient damping term to the VSM scheme,  which is related to the active-power injection of the device. Although the work focuses on transient stability (angle stability against large disturbances), the TDM damps effectively power oscillations. In \cite{amenedo21}, POD controllers in VSM-based GFOR converters are proposed using the estimated frequency at the connection point as the input signal. The work analysed POD controllers related to active- and reactive-power injections of the GFOR (POD-P and POD-Q controllers, respectively) and they were tested in a two-area system through small-signal analysis. The work in \cite{Jankovic_phd2024} proposed POD-P and POD-Q controllers for GFOR power converters also using an estimation the estimated frequency at the connection point as input signal, and also testing the proposed POD controllers in a two-area multi-machine system. In \cite{utility_based}, a supplementary POD-P  controller for oscillator-based GFOR was  proposed and tested in the IEEE 12-bus system through EMT time-domain simulations. The proposed POD-P controller used as input signal the speed deviation of a nearby generator. Reference \cite{Baruwa2021} proposed a PSS for GFOR power converters using VSM scheme, proving its effectiveness in a multi-machine/multi-converter system. The proposed PSS uses as input signal the active-power injection of the device, and it uses as output signal a supplementary voltage set-point for the GFOR. Since the control actions are linked to reactive-power injection, the PSS proposed in \cite{Baruwa2021} can be classified into POD-Q controllers.  

Since GFOR power converters mimic the behaviour of SGs, the frequency imposed by each GFOR contains very useful information about their oscillatory behaviour. Hence, it could be used as input signal of POD controllers, even though the frequency of a GFOR converter is the output of a control algorithm, and not a physical measurement. This approach offers clear implementation advantages, as the input signal is an internal variable of the controller, eliminating the need for an external measurement system, which could require additional filtering or introduce delays. To the best of the authors' knowledge, the use of POD controllers in GFORs using this type of input signals has yet to be studied. 

For example, POD controllers for GFOR power converters proposed in previous work use as input signals the frequency measured at the connection point \cite{amenedo21,Jankovic_phd2024}, a frequency measurement of a generator of the system \cite{utility_based} or the active-power injection measured at the connection point \cite{Baruwa2021}. The work in \cite{tdm_pod} indeed proposes the use of the frequency imposed by the GFOR to add a damping term. However, the work in \cite{tdm_pod} only focuses on the active-power term and the controller does not use lead/lag filters. Hence, there is room for further improvement in terms of power oscillation damping. Furthermore, the performance of POD controllers in GFOR converters in large-scale power systems with penetration of converter-interfaced generation remains relatively unexplored, since previous work only address small test systems \cite{tdm_pod,amenedo21,Jankovic_phd2024,Baruwa2021}. Another important aspect is the design of POD controllers. In previous work \cite{tdm_pod,amenedo21,Jankovic_phd2024,Baruwa2021} different design methods for POD controllers in GFOR converters were used, with the common characteristic that they were designed for a specific test system. In practice, the owners of a power plant to be commissioned often have limited information available about the large-scale power system. Therefore, it is of great interest to investigate design methods of POD controllers in GFORs using small synthetic systems, but showing a robust behaviour when they are implemented in large-scale power systems. Finally, the power-oscillation damping capability of the active-power/frequency (P-f) and reactive-power/voltage (Q-V) droop-based control structure in GFOR has yet to be studied in detail.

Along these lines, this paper investigates the performance of P-f/Q-V droop-based GFOR power converters in the damping of electromechanical oscillations, and it proposes POD-P and POD-Q controllers for this type of systems using the frequency imposed by the GFOR as input signal, which has remarkable practical implementation. Design of POD controllers for GFOR is carried out using eigenvalue-sensitivity methods using a small synthetic system. These methods were previously applied to PSSs in SGs~\cite{Rouco_2001,renedoPSS} and POD controllers in GFLs~\cite{renedoDesignPOD}, but they have not been applied to POD controllers in GFORs in previous work. For this purpose, two case studies are conducted. The first one examines a two-area system, providing fundamental insights into the effects of incorporating a droop-based GFOR and a supplementary POD controller in its active and reactive power injections (POD-P and POD-Q).  The design of POD controllers is carried out using this small two-area test system. In the second case study, the analysis is extended to the IEEE 118 bus system. The study is performed through small-signal analysis in Matlab, and time-domain results from EMT simulations in PSCAD are also provided. 

The contributions of this paper are as follows:
\begin{itemize}
    \item  Proposal of POD-P and POD-Q controllers in P-f/Q-V droop-based GFOR converters using the frequency imposed by the GFOR as the input signal, which eliminates the need for additional frequency measurements.
    \item  Design of POD-P and POD-Q controllers in GFOR power converters, using eigenvalue-sensitivity methods and a synthetic test system, with application to power systems when limited information is available. \color{black}
    \item A comparative analysis of the performance of droop-based GFOR and supplementary POD-P and POD-Q controls in damping  of electromechanical oscillations, conducted through small-signal analysis. 
    \item An assessment of POD-P and POD-Q control in GFOR within a large-scale power system with penetration of converter-interfaced generation. The same POD controller settings as in the two-area system are considered,  proving their effectiveness. \color{black}
\end{itemize}

\vspace{-1.0em}
\section{ Modelling and control of GFORs}
\label{sec:GFOR_model}

\begin{figure*}[hbt!]
\centering
\includegraphics[width=\textwidth]{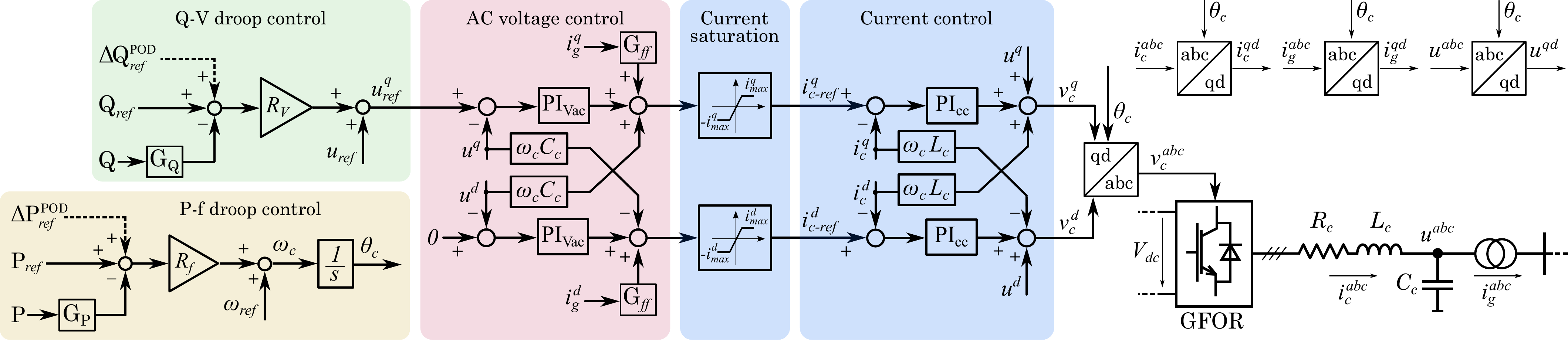}
\caption{GFOR control scheme.}
\label{fig:vsc_scheme}
\end{figure*}

\noindent The grid-forming control structure considered in this paper is shown in Fig.~\ref{fig:vsc_scheme}. The converter is represented as a controlled AC voltage source and assuming an infinite DC-bus. The GFOR converter has a  cascaded AC-voltage and current control in the $qd$-domain, using vector control. Self-synchronisation of the GFOR is achieved with an active-power (P) - frequency (f) droop control (P-f droop, for short). The GFOR also has a reactive-power (Q) - Voltage (V) droop control (Q-V droop, for short). The supplementary set points added by POD controllers will be explained in Section~\ref{sec:pod}. 

An average model of the GFOR suitable for ElectroMagnetic Transient (EMT) simulation is considered. For small-signal stability analysis, an EMT-type linearised state-space model of the power system containing GFOR power converters is considered. The modelling approach follows the guidelines of~\cite{whereisthelimit}. 

Notice that electromechanical oscillations involve slow dynamics and, therefore, electromechanical-type models (also known as Root-Mean-Square (RMS) models) of the power system and power converters could be used. Nevertheless, EMT-type models are used in this since the tools developed by the research team are also used to analyse faster interactions~\cite{whereisthelimit}. Naturally, the use of EMT models for this analysis does not compromise the fidelity of the results and it just adds further details.



\section{Power Oscillation Damping Control in GFOR}
\label{sec:pod}
\begin{figure}[!htbp]
\centering
\subfloat[POD-P control scheme]{\includegraphics[width=\columnwidth]{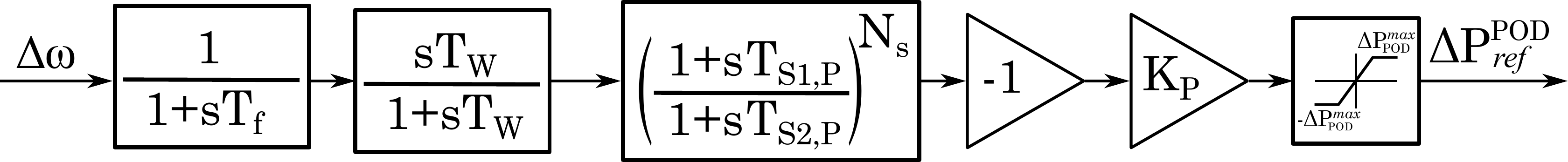}}
\vspace{0.01\textwidth}
\subfloat[POD-Q control scheme]{\includegraphics[width=\columnwidth]{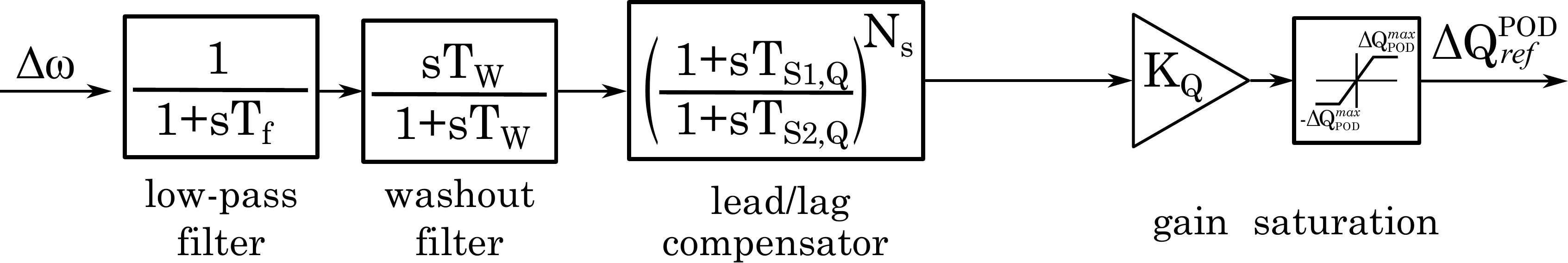}}
\vspace{0.01\textwidth}
\caption{Proposed POD controllers for GFOR converters.}
\label{fig:POD}
\end{figure}

\noindent The supplementary set-point values of the POD-P ($\subm{$\Delta \text{P}$}{POD}^\text{ref}$) and POD-Q ($\subm{$\Delta \text{Q}$}{POD}^\text{ref}$) are added to the \mbox{P-f} and \mbox{Q-V} droop control, respectively (see Fig. \ref{fig:vsc_scheme}). The block diagram of the proposed POD controller is shown in Fig. \ref{fig:POD}, which is based on the control structures used in PSSs in synchronous machines~\cite{1994power} and in POD controllers in other contexts~\cite{Rouco_2001,pod_review,pod_pv,hvdc_interarea_survey}. 


The gain ($K_P$ or $K_Q$) is adjusted to achieve the required damping ratio (typically \qtyrange{10}{15}{\percent}) of the target electromechanical modes; the washout filter acts as a high-pass filter ($\subm{T}{W}$ is in the range \qtyrange[range-phrase=--]{1}{20}{\second}), and it is used to ensure that the low-frequency oscillations are not altered by any DC component; the low-pass filter is intended for noise compensation in the input signal ($\subm{T}{f}$ is in the range \qtyrange[range-phrase=--]{0.01}{0.1}{\second}) and the lead/lag compensator provides the required phase compensation between input and output signals. POD controllers also have a saturation parameter, in order to bound their control actions. \color{black}

In this paper, frequency deviation $\Delta\omega = \omega - \omega_0$ (with \mbox{$\omega_0$ = \qty{1}{\pu}} being the converter rated frequency) is used as input signal. Notice that the frequency imposed by the GFOR is the output of a control algorithm. Hence, it has the advantage that no additional frequency measurements are needed. 

\section{Design of POD controllers for GFORs \color{black}}
\label{sec:pod_design}

POD controllers will be designed using eigenvalue-sensitivity methods~\cite{Pagola_1989}, based on the work in~\cite{Rouco_2001}. These methods have been applied to POD controllers in the context of VSC-HVDC systems in~\cite{renedoDesignPOD,MRobin_2024}. Specifically, this work follows the approach used in~\cite{MRobin_2024}, but it applies it to the design of POD controllers for GFOR power converters. In this work, POD controllers are designed using a synthetic power system, as in~\cite{renedoPSS}, and later they are tested in a large-scale power system, to prove their effectiveness. This approach is useful when limited information of the power system is available and one of the scopes of this paper is to analyse the potential of these types of methods.

A linearized model of the power system used for the design with GFOR power converters equipped with POD controllers (POD-P and POD-Q controllers)  is considered. The system will present a critical electromechanical mode $i$ (target eigenvalue):
\begin{equation}
 \label{eq:lambda0}  
 \lambda_i^0 = \sigma_i^0 \pm j \omega_i^0
\end{equation}
POD controller $j$ in the GFOR will be used to damp the target electromechanical mode $i$. The target electromechanical mode has an initial damping ratio of $\zeta_i^0$ and for the design of the POD controller a required damping ratio, $\zeta_i^d$, is specified. 

The sensitivity of the target electromechanical mode $i$ to changes in the gain of POD controller $j$ of the GFOR, $K_{j}$, is defined as~\cite{Pagola_1989}:
\begin{equation}
 \label{eq:Sij}  
 S_{ij} = \frac{\partial \lambda_i}{\partial K_{j}} 
\end{equation}
A practical way to compute the eigenvalue sensitivity is using a hybrid representation of the linearized system~\cite{Pagola_1989}.

The non-compensated (NC) sensitivity of the target electromechanical mode $i$ to changes in the gain of the non-compensated POD controller $j$ (without lead/lag compensation) is given by:
\begin{eqnarray}
 \label{eq:SijNC}  
 S_{ij}^{NC} &=& S_{ij}|_{T_{S1,j}=T_{S2,j}=0} = \frac{\partial \lambda_i}{\partial K_{j}}\bigg|_{T_{S1,j}=T_{S2,j}=0} \\ \nonumber
 &=& |S_{ij}^{NC}| \angle \varphi_{ij}^{NC} 
\end{eqnarray} 

\begin{figure}[!htbp]
\centering
\includegraphics[width=0.3\textwidth]{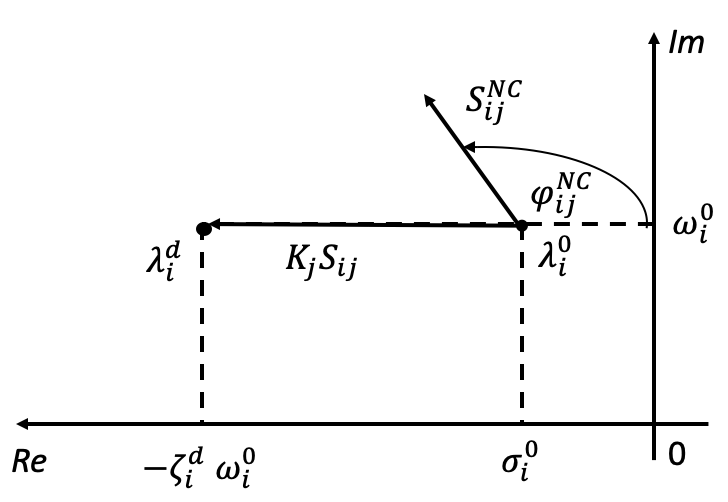}
\caption {Geometric interpretation of eigenvalue sensitivities and POD controllers.}
\label{fig:eig_sensitivity}
\end{figure}

The geometric interpretation of eigenvalue sensitivities and the effect of POD controller $j$ are illustrated in Fig.~\ref{fig:eig_sensitivity}~\cite{Rouco_2001,renedoDesignPOD}. An effective POD controller will push the target electromechanical mode $i$ to the left-hand side of the complex plane. Lead/lag filters of the POD controller are used to obtain an angle of the eigenvalue sensitivity ($S_{ij}$) close to 180º, whilst POD gain is used to achieve the required damping ratio. In this way, the target electromechanical mode (with the required damping ratio) can be estimated as:  
\begin{equation}
 \label{eq:lambdad}  
 \lambda_i^d = - \zeta_i^d \omega_i^0 \pm j \omega_i^0
\end{equation}

Notice also that $S_{ij}=\partial \lambda_i/\partial K_{j}$ is the eigenvalue sensitivity with the POD controller and with its lead/lag filters implemented.

The design of GFOR's POD-$j$ involves two steps~\cite{Rouco_2001}:
\begin{enumerate}
    \item {\em Lead/lag filters:} The objective is to obtain a phase of the eigenvalue sensitivity ($S_{ij}$) close to $180^{{\rm o}}$~\cite{Rouco_2001}. The phase of the non-compensated eigenvalue sensitivity is  $\varphi_{ij}^{NC}=\angle{(S_{ij}^{NC})} \in [-180^{{\rm o}}, 180^{{\rm o}}]$ (see Fig.~\ref{fig:eig_sensitivity}) and $a_j=T_{S2,j}/T_{S1,j}$ is the filtering ratio. Then, the latter is calculated as:
    \begin{itemize}
        \item If $\varphi_{ij}^{NC} \geq 0$, a lead compensation is required:
            \begin{equation}
        a_{j} = \frac{1 - \sin \phi_{ij}}{1 + \sin \phi_{ij}} \le 1,  \;\phi_{ij} = \frac{\pi-\varphi_{ij}^{NC}}{N_{S,j}}. \label{eq.phaselead1}
            \end{equation}

        \item If $\varphi_{ij}^{NC} < 0$, a lag compensation is required:
            \begin{equation}
        a_{j} = \frac{1 + \sin \phi_{ij}}{1 - \sin \phi_{ij}} >1, \; \phi_{ij} = \frac{\pi+\varphi_{ij}^{NC}}{N_{S,j}}. \label{eq.phaselag2}
            \end{equation}
    \end{itemize}

    Finally, parameters of the lead/lag filter ($T_{S1,j}$ and $T_{S2,j}$) are selected centering the filter at the frequency of the target electromechanical mode $i$ ($\omega_i^0$ in rad/s), as in~\cite{larsen}:
    
\begin{equation}\label{eq:leadlag_TQ1}
    T_{S1,j} = \frac{1}{\omega_i^0 \sqrt{a_{j}}}, \mbox{\space and \space} T_{S2,j}=a_{j} T_{S1,j}.
\end{equation}

    \item {\em POD gain:} Gain $K_{j}$ is calculated to obtain the required damping ratio of the target electromechanical mode $i$: 
    \begin{equation}
 \label{eq:KSj}  
 K_{j} = \frac{|\lambda_i^d-\lambda_i^0|}{|\frac{\partial \lambda_i}{\partial K_{j}} |}
\end{equation}
with $K_{j} \in [K_{j}^{min}, K_{j}^{max}]$. Notice that, since the design will be carried out in a synthetic system, a minimum gain $K_{j}^{min}$ is considered in the design process to ensure that the POD controller is effective in a larger power system. 
\end{enumerate}


This work approximates the non-compensated eigenvalue sensitivity used for the design numerically, using a finite-difference calculation, as proposed in \cite{MRobin_2024}, due to its practical implementation:
\begin{equation}
 \label{eq:hatSijNC}  
 \hat S_{ij}^{NC} = \frac{\lambda_i^{NC}-\lambda_i^0}{\Delta K_{j}} \approx \frac{\partial \lambda_i}{\partial K_{j}}\bigg|_{T_{S1,j}=T_{S2,j}=0}
\end{equation}
where $\lambda_i^{NC}$ is the electromechanical mode obtained with the non-compensated POD controller (e. g., with $T_{S1,j}=T_{S2,j}=0$ s) and using a small gain value $\Delta K_{j}$.

\section{Case study 1: Two-area system}\label{sec.case1_results}
\begin{figure}[H]
\centering
\includegraphics[width=0.8\columnwidth]{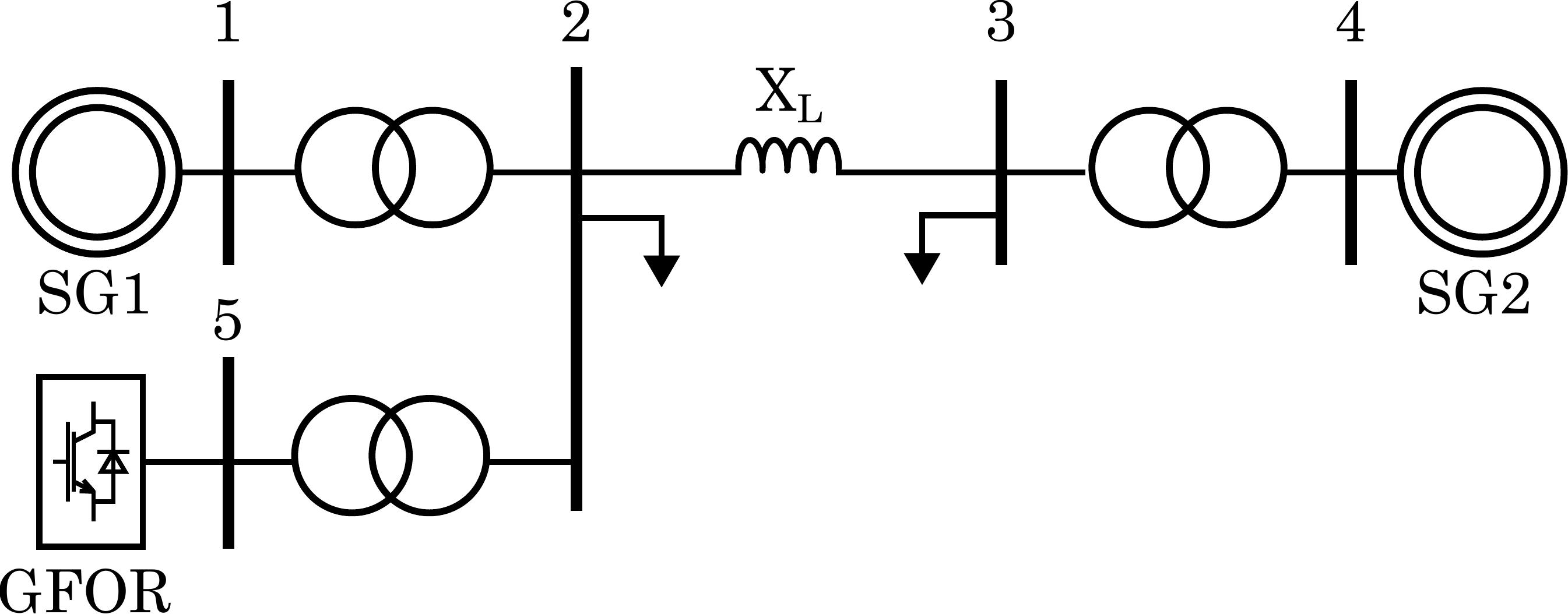}
\caption{Two-area synthetic system \cite{nts}.}
\label{fig_nts_sys}
\end{figure}

\noindent In this section, the damping capability of GFOR and the POD control described in the previous section are assessed in the two-machine system shown in Fig. \ref{fig_nts_sys}. This synthetic system is the benchmark established by the Spanish Transmission System Operator (TSO) to evaluate the requirements for generator units to connect to the network \cite{nts}, which is based on the two-area synthetic test system and methodology proposed in \cite{renedoPSS}. The purpose of this synthetic two-area test system is twofold:
\begin{enumerate}
    \item It will be used for the design of POD controllers in GFORs, which will be used in a large-scale power system in Section \ref{sec.case2_results}. 
    \item For fundamental analysis of the performance of the proposed POD controllers for GFORs.
\end{enumerate}
\color{black}

The synthetic system consists of two synchronous generators, SG1 and SG2, of \qty{1500}{\mega\voltampere} and \qty{5000}{\mega\voltampere} rated power, respectively. A GFOR of the same rated power as SG1 (\qty{1500}{\mega\voltampere}) is connected to bus 2. The reactance of the step-up transformers is \qty{0.15}{\pu} in the \unit{\pu}-base of each element. Loads in buses \mbox{2-3} are represented as constant impedance loads. The system is operating with \qty{100}{\mega\watt} flowing through the line \mbox{2-3}. In the operating point, SG1 and GFOR are injecting an active power of $P_{SG1}=P_{GFOR}=\qty{1350}{\mega\watt}$. The rest of operating point details are found in \cite{nts}. It should be noticed that PSSs of SG1 has been removed, in order to obtain a more critical case. 


The modeling of the SGs and their parameters are described in detail in \cite{nts}. The GFOR implementation has been described in Section \ref{sec:pod}, and values of the converter parameters are provided in Table \ref{tab:gfor_param} of Appendix. 

The methodology used in this case study to analyze inter-area oscillations within a specified frequency range follows the approach proposed in \cite{renedoPSS}. Specifically, the oscillation frequency and damping of the electromechanical mode are modified based on the value of the line reactance $\subm{X}{L}$, which is considered within the range of \qtyrange{0.01}{0.6}{\pu} (using a system \unit{\pu}-base of \qty{100}{\mega\voltampere} and \qty{230}{\kilo\volt}). High values of $\subm{X}{L}$ reproduce inter-area oscillations, whereas low values of $\subm{X}{L}$ correspond to local oscillations of the SGs. The goal is to increase the damping ratio of the electromechanical mode at several frequencies without compromising performance across the entire frequency range.

\vspace{-1.0em}
\subsection{Design of the POD controllers}\label{sec.case1_results_POD_design}
\noindent Case study 1 (synthetic test system of Fig. \ref{fig_nts_sys}) is used to design the POD controllers. POD-P and POD-Q controllers of the GFOR connected to bus 5 (Fig. \ref{fig_nts_sys}) are designed separately, following the methodology presented in Section~\ref{sec:pod_design}. A single operating point is used for the design ($\subm{X}{L} = \qty{0.1}{\pu}$, which corresponds to an oscillation-frequency about \qty{0.58}{\hertz}), but the robustness of the designed POD controllers will be assessed in the full frequency range of the electromechanical oscillations ($\subm{X}{L}$ from \qtyrange{0.01}{0.6}{\pu}). The obtained POD controllers in GFORs will be tested in a large-scale power system in Section~\ref{sec.case2_results} (Case study 2), which has critical electromechanical modes within frequency range \qtyrange[range-phrase=--]{0.67}{0.9}{\hertz}.

For the design, a \qty{10}{\percent} increment of damping ratio ($\Delta \zeta_d$) with respect to the case with GFOR but without POD controllers, is specified. POD-P and POD-Q controllers have the structure described in Fig. \ref{fig:POD}. Parameters $T_f=\qty{0.1}{\second}$, $T_W=\qty{5}{\second}$, $N_S=\num{2}$ and $\Delta P_{POD}^{max}=\Delta Q_{POD}^{max}=\qty{0.2}{\pu}$ are set before the design. The design process calculates the gains and lead/lag filters of POD-P and POD-Q controllers. Gain values are bounded withing a range of $K_P, K_Q \in [200, 400]$ \unit{\pu}. As explained in Section~\ref{sec:pod_design}, a minimum value for each gain is set to ensure that the controllers are effective when considering larger power systems.

Table \ref{tab:gfor_POD_param} shows the parameters of POD-P and POD-Q controllers obtained with the design method. Next section will analyse the results obtained with the proposed POD controllers for GFORs. Notice that gain values of POD-P and POD-Q controllers correspond to their minimum value used during the design process ($K_P= K_Q=\qty{200}{\pu}$), which means that lower gain values are required to obtain the specified damping ratio of the electromechanical mode in the synthetic system used.

\vspace{-1.0em}
\begin{table}[!htbp]
\renewcommand{\arraystretch}{0.7}
\centering
\captionsetup{justification=centering, labelsep=newline, textfont=footnotesize}
\caption{Parameters of POD controllers in GFORs.}
\label{tab:gfor_POD_param}
\begin{tabular}{lcccccc}
\toprule
        & $\subm{T}{f}$ [\unit{\second}] & $\subm{T}{W}$ [\unit{\second}]  & $\subm{T}{S1}$ [\unit{\second}] & $\subm{T}{S2}$ [\unit{\second}] & $\subm{N}{S}$   & $\text{K}$ \\
\midrule
POD-P   & \num{0.1}  & \num{5}  & \num{0.20} & \num{0.37} & \num{2}         & \num{200} \\
POD-Q   & \num{0.1}  & \num{5}  & \num{0.26} & \num{0.29} & \num{2}         & \num{200} \\
\bottomrule
\multicolumn{7}{l}{\footnotesize pu base: converter rated values}
\end{tabular}
\end{table}

\vspace{-1.0em}
\subsection{Small-signal analysis}\label{sec.case1_results_SSA}
\begin{figure}[hbt!] 
\centering
\includegraphics[width=\columnwidth]{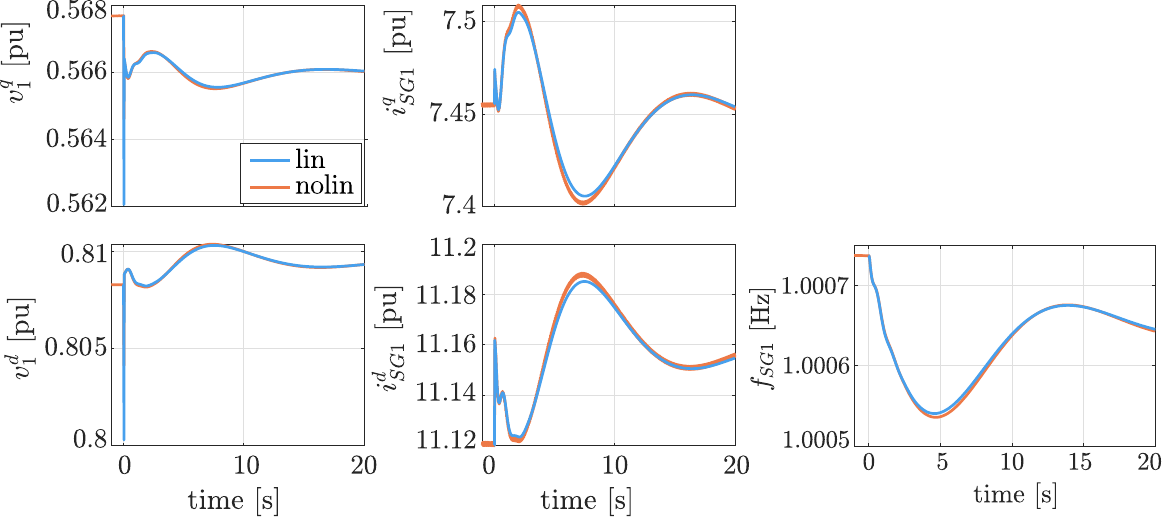}
\caption{Nonlinear vs. linear model of the system in Fig. \ref{fig_nts_sys}, for Base case, when a load increase of \qty{1}{\percent} is applied at bus 2. Showing $qd$ voltages, currents, and frequency of SG1.}
\label{fig:validation_NTS}
\end{figure}

\noindent This section presents the small-signal results of the first case study. Fig. \ref{fig:validation_NTS} illustrates the validation of the state-space model of the system, implemented in Matlab, with respect with a non-linear EMT model in PSCAD/EMTDC software. Results show good agreement. 

In order to study the effect of GFOR and POD control on the damping of the electromechanical mode, the following five scenarios are analyzed and compared: 
\begin{enumerate}[leftmargin=*]
    \item  \textbf{Base (only SG):} only synchronous generators SG1 and SG2 are connected. This scenario is considered as the base case for the subsequent comparisons.
    \item  \textbf{SG + GFOR:} \mbox{GFOR without POD is connected at bus 2.}
    \item  \textbf{SG + GFOR POD-P:} POD-P is activated in GFOR. 
    \item  \textbf{SG + GFOR POD-Q:} POD-Q is activated in GFOR.
    \item  \textbf{SG + GFOR POD-PQ:} POD-P and POD-Q are activated in GFOR.
\end{enumerate}


First, the results for the scenario used to design the POD controllers in the GFOR are analyzed ($\subm{X}{L}$ = \qty{0.1}{\pu}). The results are presented in Table~\ref{tab:nts_damp}. In the Base case (only SG), the electromechanical mode has a damping ratio of $\zeta=\qty{-1.63}{\percent}$ (unstable) and a frequency of $f=\qty{0.58}{\hertz}$. When the GFOR is included (SG + GFOR), the damping ratio increases to $\zeta=\qty{10.3}{\percent}$, due to the inherent damping capability of the droop-controlled GFOR power converters. When POD-P and POD-Q controllers are implemented in the GFOR, with the design of Subsection~\ref{sec.case1_results_POD_design}, significant improvements are obtained. The damping ratio obtained with POD-P controller, POD-Q controller and with both simultaneously (POD-PQ) are $\zeta=\qty{22.4}{\percent}$, $\zeta=\qty{24.4}{\percent}$ and $\zeta=\qty{24.0}{\percent}$, respectively.  

\begin{table}[h!]
\addtolength{\tabcolsep}{-1.5pt}
\renewcommand{\arraystretch}{0.9}
\centering
\captionsetup{justification=centering, labelsep=newline, textfont=footnotesize}
\caption{Characteristics of electromechanical mode when $\subm{X}{L}=\qty{0.1}{\pu}$.}
\label{tab:nts_damp}
\begin{tabular}{lcccc}

\toprule
Case    & Trajectory & Mode &  $f$ [\unit{\hertz}] & $\xi$ [\%]   \\
\midrule
Base (only SG)         & (a) & \complexnum[output-complex-root = j,complex-root-position = before-number]{0.06 \pm j3.62}    & \num{0.58}    & \num{-1.63}   \\
SG + GFOR            & (b) & \complexnum[output-complex-root = j,complex-root-position = before-number]{-0.38 \pm j3.66}     & \num{0.58}    & \num{10.3}   \\
SG + GFOR POD-P      & (d) & \complexnum[output-complex-root = j,complex-root-position = before-number]{-1.10 \pm j4.80}     & \num{0.76}    & \num{22.4}   \\
SG + GFOR POD-Q      & (c) & \complexnum[output-complex-root = j,complex-root-position = before-number]{-0.91 \pm j3.62}    & \num{0.58}    & \num{24.4}   \\   
SG + GFOR POD-PQ     & (e) & \complexnum[output-complex-root = j,complex-root-position = before-number]{-1.26 \pm j5.09}     & \num{0.81}    & \num{24.0}   \\   
\bottomrule
\end{tabular}
\end{table}



\begin{figure}[h!]
\centering
\includegraphics[width=0.95\columnwidth]{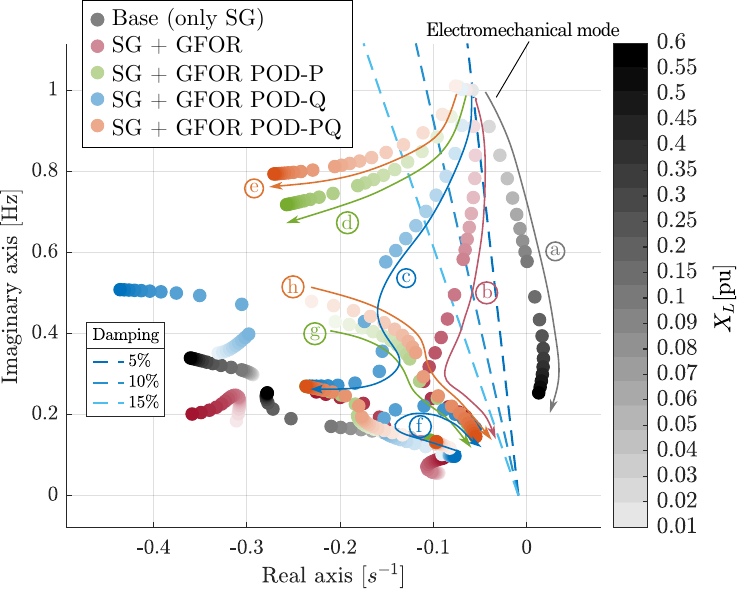}
\caption{Trajectories of the low-frequency eigenvalues for each scenario.} 
\label{fig_nts_modes_2}
\end{figure}

The impact of GFOR power converters and POD controllers is now analyzed for electromechanical modes of different oscillation frequencies. Fig. \ref{fig_nts_modes_2} shows the low-frequency eigenvalues of the system for the five mentioned scenarios when $\subm{X}{L}$ is varied. Notice that the imaginary part of the eigenvalues is shown in [\unit{\hertz}] instead that in [rad/s], in order to identify directly the oscillation frequency of each mode. To facilitate the analysis, the electromechanical mode trajectories are identified with letters from (a) to (h). Table \ref{tab:nts_damp} details the oscillation frequency and damping at the design point ($\subm{X}{L}$ = \qty{0.1}{\pu}) for each scenario.

Results from Fig. \ref{fig_nts_modes_2} show that as $\subm{X}{L}$ is increased, the oscillation frequency is reduced, corresponding to an inter-area interaction for high $\subm{X}{L}$ values. This is also confirmed by a participation factor analysis \cite{pf,Pagola_1989}, considering the relative participation of the state variables $\omega_{\text{SG1}}$ and $\omega_{\text{SG2}}$ (SGs frequency) to the electromechanical mode. Note that the specific results of this participation factor analysis are not included, due to the lack of room.  


In the Base case, which corresponds to the trajectory (a) in Fig.~\ref{fig_nts_modes_2}, the electromechanical mode is unstable for most of values of the line reactance $\subm{X}{L}$, since its real part is positive. Notice that this occurs due to the fact that SG1 does not have a PSS (which was deactivated for illustrative purposes). In the case with GFOR but without POD controllers, (trajectory (b) in Fig.~\ref{fig_nts_modes_2}) the damping ratio of the electromechanical mode increases in all the frequency range, in comparison with the case without GFOR, due to its inherent damping capability. This can be observed in Fig.~\ref{fig_nts_modes_2}, where, due to the presence of the GFOR, the eigenvalue trajectory is moved to the left-hand side of the complex plane. 

Fig.~\ref{fig_nts_modes_2} also illustrates the impact of POD controllers in the GFOR on the trajectories of the electromechanical mode in the complex plane as $\subm{X}{L}$ changes: POD-P (trajectories (d)-(g)), POD-Q (trajectories (c)-(f)) and POD-PQ (trajectories (e)-(h)). In the case of POD-P controller, the trajectory of the electromechanical mode is split into two paths: for high oscillation frequencies (low values of $\subm{X}{L}$), the electromechanical mode follows the trajectory marked with a (d) in Fig.~\ref{fig_nts_modes_2}, while for low oscillation frequencies (high values of $\subm{X}{L}$), the electromechanical mode follows the trajectory marked with a (g) in Fig.~\ref{fig_nts_modes_2}. This behavior occurs because POD controllers naturally influence the electromechanical mode, and participation factors of the modes of the system can change as $\subm{X}{L}$ changes. In the cases of POD-Q and POD-PQ controllers in the GFOR, the electromechanical mode follows a similar pattern (trajectories (c)-(f) and (e)-(h) of Fig.~\ref{fig_nts_modes_2}, respectively). Results show that POD controllers in the GFOR move the electromechanical mode to the left-hand side of the complex plane across the entire frequency range, significantly increasing its damping ratio.


Overall, both Fig.~\ref{fig_nts_modes_2} and Table~\ref{tab:nts_damp} show that GFOR without POD exhibits an inherent damping capability which increases the damping ratio of electromechanical modes. Nevertheless, significant improvements can be achieved with POD controllers implemented in the GFOR. Results show that, in general, POD-P controller achieves the highest damping ratio of the electromechanical mode across the entire range of $\subm{X}{L}$ tested values, particularly at the local oscillation region. While POD-Q controller is also effective, in general it provides less damping of the electromechanical mode compared to POD-P. Nevertheless, POD-Q controller has the advantage that its control actions are not linked to the primary energy source of the GFOR. Results also show that further improvements can be achieved by implementing simultaneously POD-P and POD-Q controllers (POD-PQ case). Finally, it is worth to highlight that, although the design of POD controllers was performed for a single operating point $\subm{X}{L}$ = \qty{0.1}{\pu}, they present a robust behaviour, since they are effective to damp electromechanical oscillations in the whole frequency range analyzed (line reactances $\subm{X}{L}$ from \qtyrange{0.01}{0.6}{\pu}).

\vspace{-1.0em}
\subsection{Non-linear time-domain simulation}
\noindent This section presents non-linear time-domain results from EMT simulations in PSCAD/EMTDC for the case with $\subm{X}{L}$ = \qty{0.1}{\pu}  when a load increase of \qty{1}{\percent} is applied at bus 2 (e.g., a small disturbance). \color{black}

\begin{figure}[h!]
\centering
\includegraphics[width=\columnwidth]{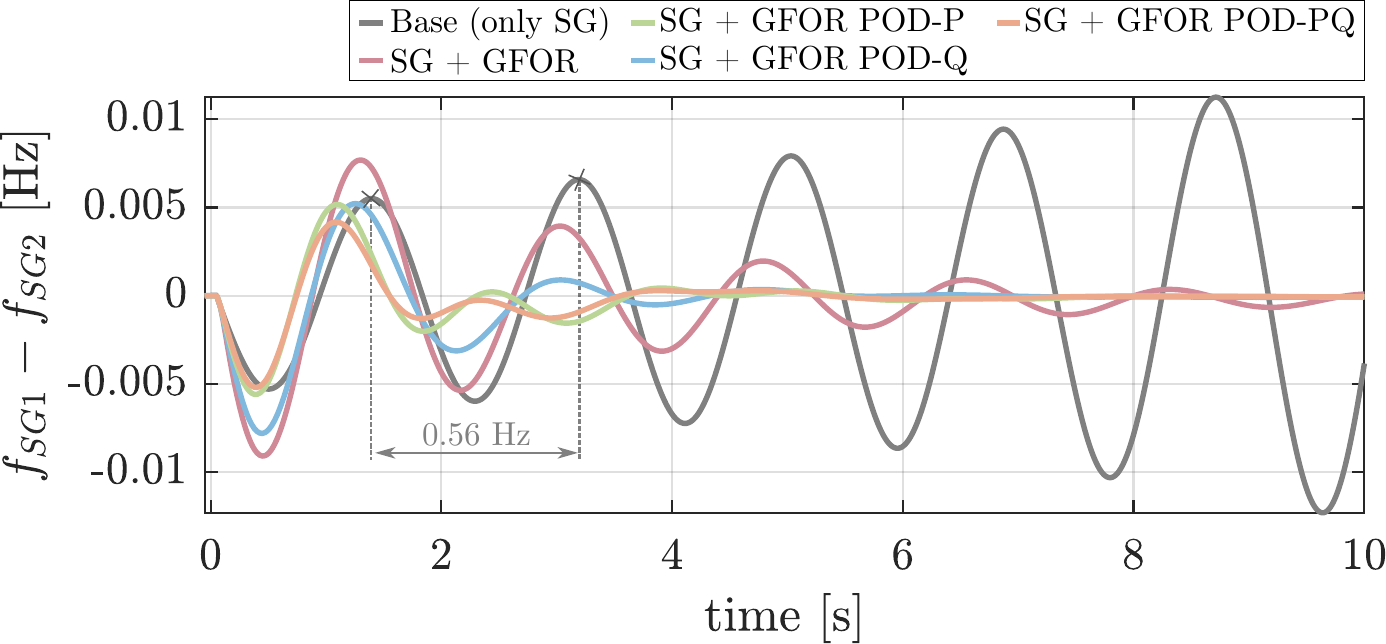}
\caption{Frequency difference between SG1 and SG2 ($\subm{X}{L} = \qty{0.1}{\pu}$).}
\label{fig_nts_freq}
\end{figure}

In Fig.~\ref{fig_nts_freq}, the frequency difference between SG1 and SG2 is shown for the five scenarios considered. Results confirm consistency with the ones presented in Subsection~\ref{sec.case1_results_SSA} (see Table~\ref{tab:nts_damp}). The Base case is unstable and, therefore, the electromechanical oscillation is increasing with time. In the presence of a droop-controller GFOR power converter (SG+GFOR) the damping ratio of the electromechanical mode increases in comparison with the Base case. Results show that when using POD-P, POD-Q and POD-PQ controllers in the GFOR converter, the electromechanical mode has high-damped dynamic response. 

\begin{figure}[h!]
\centering
\includegraphics[width=\columnwidth]{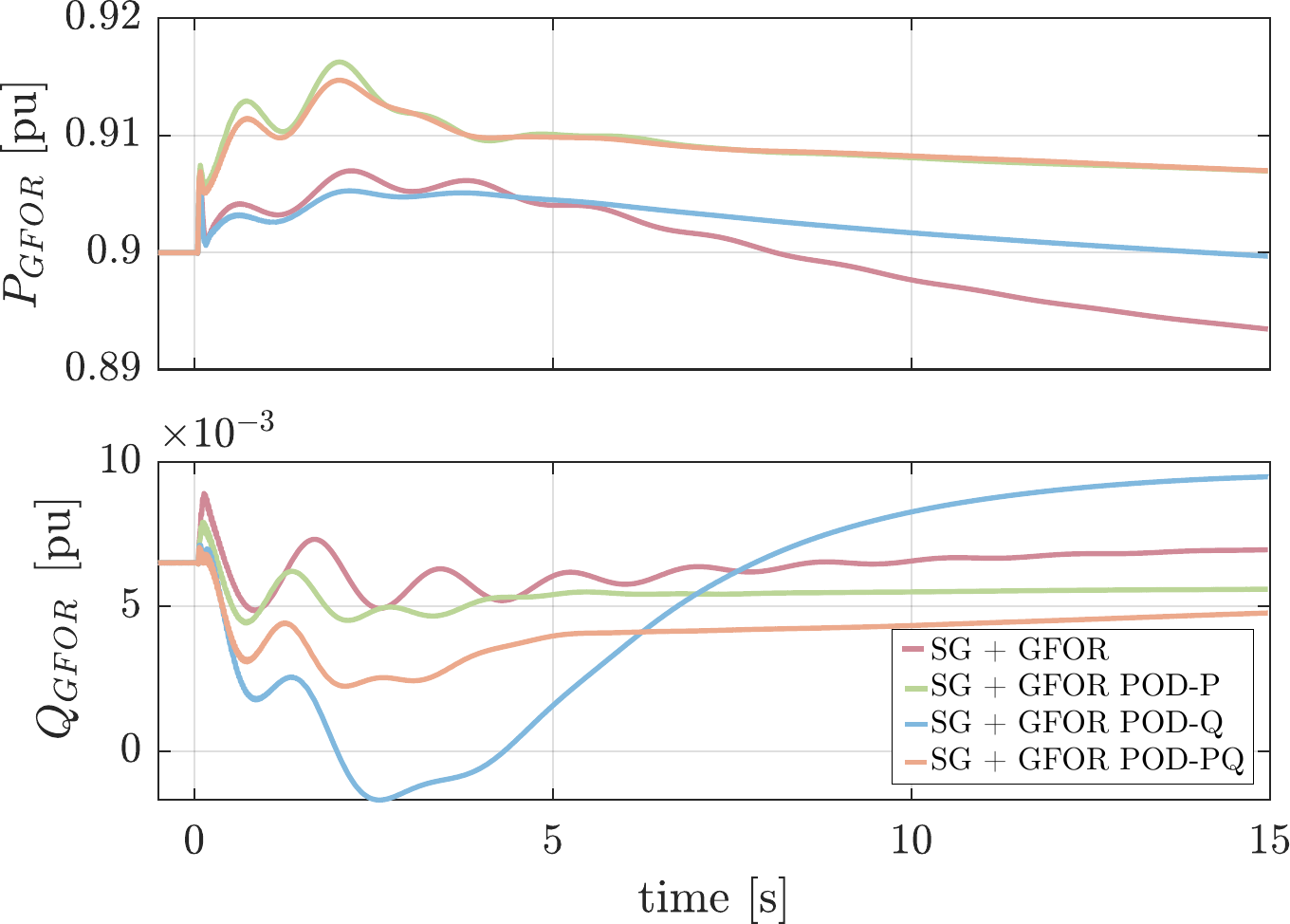}
\caption{Active and reactive power injection of GFOR ($\subm{X}{L} = \qty{0.1}{\pu}$).}
\label{fig_nts_PQ}
\end{figure}

In Fig.~\ref{fig_nts_PQ}, the active and reactive power injection of GFOR is shown for the four scenarios involving the converter. In the case of GFOR without POD controllers, changes on the converter P/Q injections are due to the P-f and Q-V droop controllers, which react to the load change. In the case of GFOR with POD controllers changes on the converter P/Q injections are due to both, the effect of P-f/Q-V droop controllers and P/Q modulation due to POD controllers. 

\begin{figure}[h]
\centering
\includegraphics[width=\columnwidth]{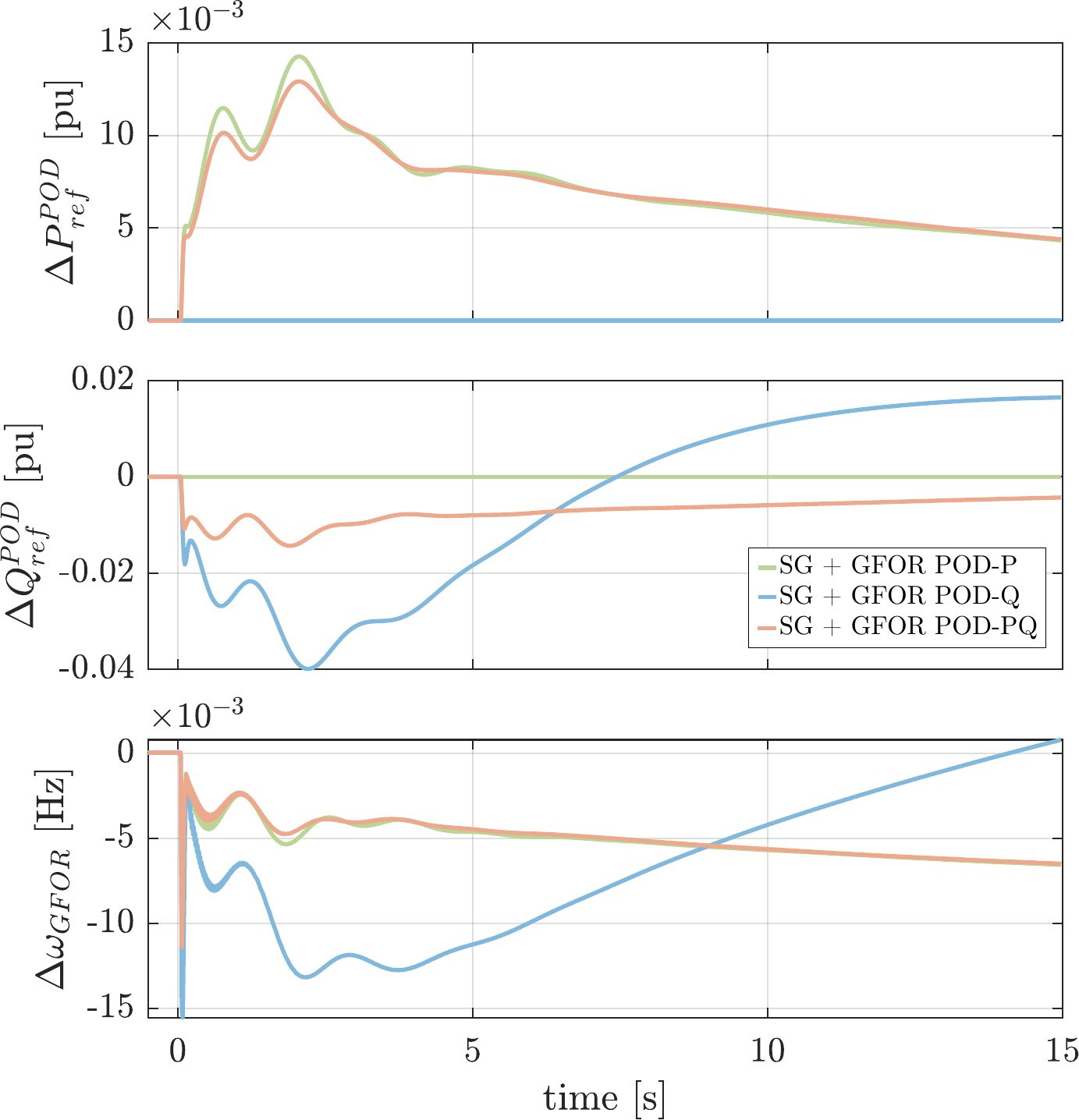}
\caption{Input (GFOR frequency deviation) and Output (Supplementary active and reactive power references) signals from POD control ($\subm{X}{L} = \qty{0.1}{\pu}$).}
\label{fig_nts_POD_PQ_refs}
\end{figure}

To further understand the effect of POD controllers, Fig.~\ref{fig_nts_POD_PQ_refs} shows the frequency imposed by the GFOR (input signal of POD controllers in \unit{\pu}) and supplementary set-point values for P/Q injections provided by POD controllers (written in \unit{\pu} with respect to GFOR's nominal apparent power) (see Fig.~\ref{fig:POD}). Control actions of POD controllers change accordingly to the oscillations seen in the input signal (frequency deviation) and they damp successfully the electromechanical oscillation, as shown in Fig.~\ref{fig_nts_freq}.

\newpage

\section{Case study 2: IEEE 118-bus system}\label{sec.case2_results}
\noindent In this section, the damping capability of GFORs and the performance of POD-P, POD-Q and POD-PQ controllers in GFOR (previously described in Section~\ref{sec:pod}) are  assessed in a large-scale system with both synchronous and converter-interfaced generation. The objective is to validate whether the insights obtained from the first case study, conducted on a synthetic two-machine system, are still valid when extrapolated to larger and more realistic systems. The design of POD controllers is the one obtained for the synthetic system (Table~\ref{tab:gfor_POD_param} of Section~\ref{sec.case1_results_POD_design}). Precisely, one of the objectives of this paper is to investigate design of POD controllers in GFORs using synthetic test systems, which is remarkably useful in practice when limited information of the power system is available.  \color{black}

\begin{figure}[b]
\centering
\includegraphics[width=\columnwidth]{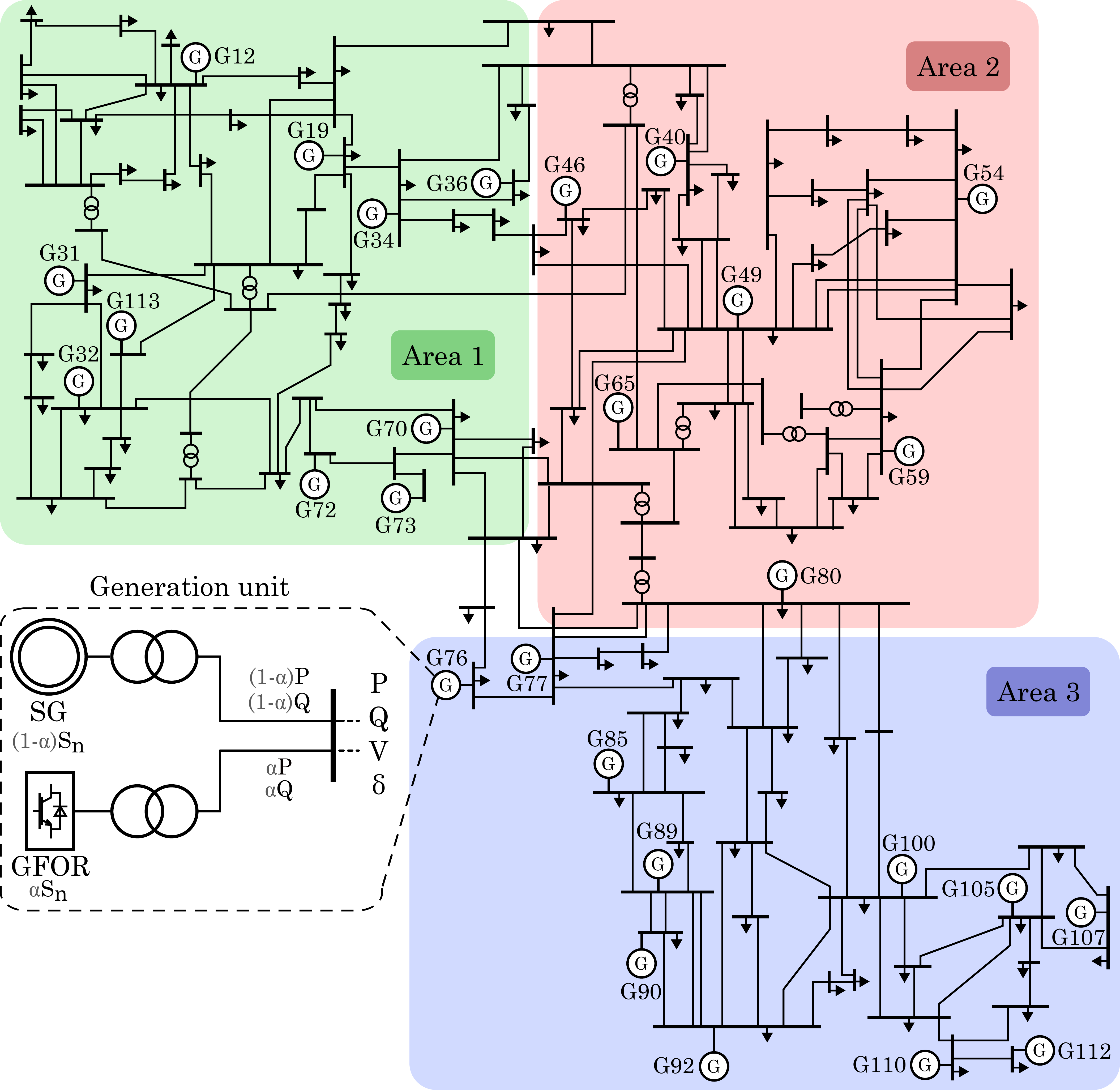}
\caption{IEEE 118-bus system with each generator unit comprising an SG and GFOR, with adjustable penetration levels in the total power injection.}
\label{fig_118_sys}
\end{figure}

This second case study is performed using the IEEE 118-bus system benchmark provided by PSCAD, which consists of transmission lines, loads, and 28 generator units~\cite{PSCADSpec}. Further power stability studies using this test system in the presence of converter-interfaced generation can be found in \cite{Ndreko_2018,Castro_2023}. In this work, the benchmark system has been modified so that each generator unit includes both a SG and a GFOR, as shown in Fig.~\ref{fig_118_sys}. The contribution of each element to the total nominal apparent power and total power injection of the generator unit is weighted by the parameter $\alpha$, while maintaining the total rating and power injection of the generator connected at each bus in the original case. Also, the loads are represented as constant impedances. Transmission system parameters and power-flow data can be found in \cite{PSCADSpec}. Parameters of the SGs are given in Table \ref{tab:sg_param} of the Appendix, and GFORs are implemented using the same control structure and parameters of Case study 1. Notice that SGs are not equipped with PSSs, in order to consider critical cases for electromechanical oscillations and focus the attention on the performance of POD controllers in the GFORs.  

The same five scenarios from the previous case study are considered. In particular, Base (only SG) corresponds to 100\% of synchronous generation ($\alpha$ = \num{0}); and the cases involving GFOR are analyzed with a 75\% SG -- 25\% GFOR generation mix ($\alpha$ = \num{0.25}), uniformly distributed in each generation unit. 

\vspace{-1.0em}
\subsection{Small-signal analysis}
\begin{figure}[h!]
\centering
\includegraphics[width=0.95\columnwidth]{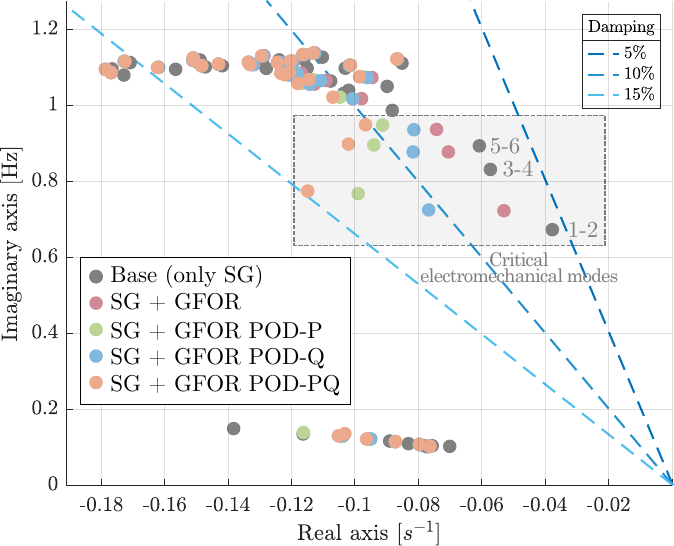}
\caption{IEEE 118-bus system eigenvalues within the \qtyrange{0.1}{2}{\hertz} frequency range for the five considered scenarios.}
\label{fig_118_modes}
\end{figure}

\noindent This section presents the small-signal results of the second case study. Fig.~\ref{fig_118_modes} shows the oscillatory modes found within the \qtyrange{0.1}{1.2}{\hertz} frequency range for the five scenarios. Again, the imaginary part of the eigenvalues is shown in [\unit{\hertz}], in order to directly identify the oscillation frequency of each mode. It is observed that there are modes in the region of damping from \num{5} to \qty{10}{\percent}, while no modes exhibit less than \qty{5}{\percent} damping. Three critical electromechanical modes have been identified in the Base case: mode 1-2, 3-4 and 5-6. \color{black}

\begin{figure}[ht!]
\centering
\includegraphics[width=\columnwidth]{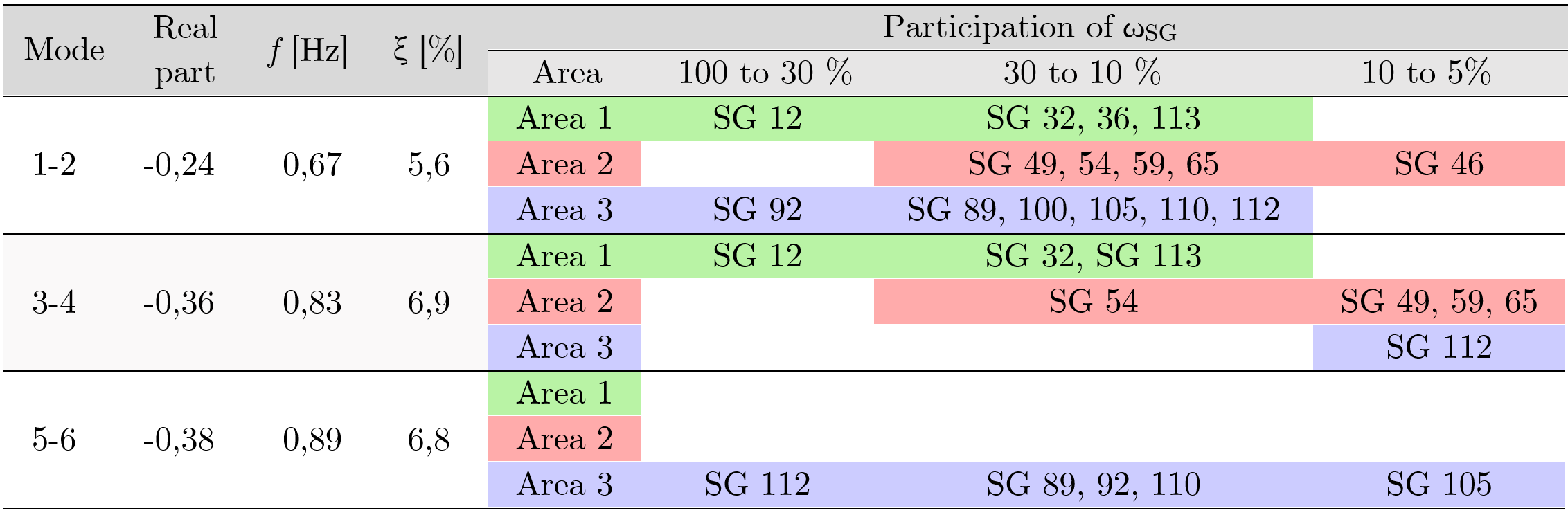}
\caption{Frequency, damping, and participation of state SG frequency ($\omega_{\text{SG}}$) for the three modes from Fig. \ref{fig_118_modes} with the lowest damping in the Base case.} 
\label{fig_118_SG}
\end{figure}

Fig. \ref{fig_118_SG} offers a detailed view of the three modes with less damping in the Base case. The SGs and areas involved in each mode have been identified through participation factor analysis. It is observed in Fig. \ref{fig_118_SG} that modes \mbox{1-2} and \mbox{3-4} correspond to electromechanical interactions between SGs from different areas, with the former exhibiting the lowest damping ratio. In particular, synchronous generators \mbox{SG-12} and \mbox{SG-92}, located in \mbox{Area 1} and \mbox{Area 3}, respectively, have the highest participation in this mode. 

In order to assess the effect of GFOR and POD controllers, \mbox{Table \ref{tab:118_damp}} provides the values of frequency and damping ratio of these three low-damped modes for the five scenarios. These modes are also indicated with a box in Fig.~\ref{fig_118_modes}. It is confirmed from both Fig. \ref{fig_118_modes} and Table \ref{tab:118_damp} that the replacement of synchronous generation by GFOR inherently contributes to the damping of the electromechanical modes, particularly in the \qtyrange{0.6}{1}{\hertz} frequency range. 

Regarding the effect of POD controllers, results of Fig.~\ref{fig_118_modes} and values of damping ratio of the critical electromechanical modes in Table~\ref{tab:118_damp} clearly indicate that results are improved when POD controllers are included. In particular, POD-P achieves a higher damping than \mbox{POD-Q}. Moreover, under POD-P, only one electromechanical mode remains within the 10\% damping range, and \mbox{mode 1-2} is remarkably shifted to the left. Results are improved further when POD-P and POD-Q controllers are implemented simultaneously. Notice also that no adverse effect of POD controllers on other modes was observed. These results align with the observations derived from the first case study. 

\begin{table}[h!]
\addtolength{\tabcolsep}{-1.5pt}
\renewcommand{\arraystretch}{0.85}
\centering
\captionsetup{justification=centering, labelsep=newline, textfont=footnotesize}
\caption{Characteristics of electromechanical modes in Fig. \ref{fig_118_modes}.}
\label{tab:118_damp}
\begin{tabular}{lcccccc}
\toprule
 & \multicolumn{2}{c}{Mode 1-2} & \multicolumn{2}{c}{Mode 3-4} & \multicolumn{2}{c}{Mode 5-6} \\
\cmidrule(r){2-3} \cmidrule(r){4-5} \cmidrule(r){6-7}
Case    & $f$ [\unit{\hertz}] & $\xi$ [\%] & $f$ [\unit{\hertz}] & $\xi$ [\%] & $f$ [\unit{\hertz}] & $\xi$ [\%]   \\
\cmidrule(r){1-1} \cmidrule(r){2-3} \cmidrule(r){4-5} \cmidrule(r){6-7}
Base (only SG)         & \num{0.67}    & \num{5.6}   & \num{0.83} &  \num{6.9}   & \num{0.89} &  \num{6.8}\\
SG + GFOR              & \num{0.72}    & \num{7.2}   & \num{0.88} &  \num{8.1}   & \num{0.94} &  \num{7.9}\\
SG + GFOR POD-P        & \num{0.77}    & \num{12.8}   & \num{0.89} &  \num{10.5}   & \num{0.95} &  \num{9.6}\\
SG + GFOR POD-Q        & \num{0.72}    & \num{10.6}   & \num{0.86} &  \num{9.3}   & \num{0.93} &  \num{8.7}\\   
SG + GFOR POD-PQ       & \num{0.77}    & \num{14.7}   & \num{0.90} &  \num{11.3}   & \num{0.95} &  \num{10.2}\\   
\bottomrule
\end{tabular}
\end{table}

\vspace{-0.5em}
\subsection{Non-linear time-domain simulation}
\noindent This section presents non-linear time-domain results from EMT simulations in PSCAD/EMTDC. A load increase of \qty{1}{\percent} is applied at bus 12 (a small disturbance), where the synchronous generator with the highest participation in mode \mbox{1-2} (\mbox{SG-12}) is connected. 

\begin{figure}[h!]
\centering
\includegraphics[width=\columnwidth]{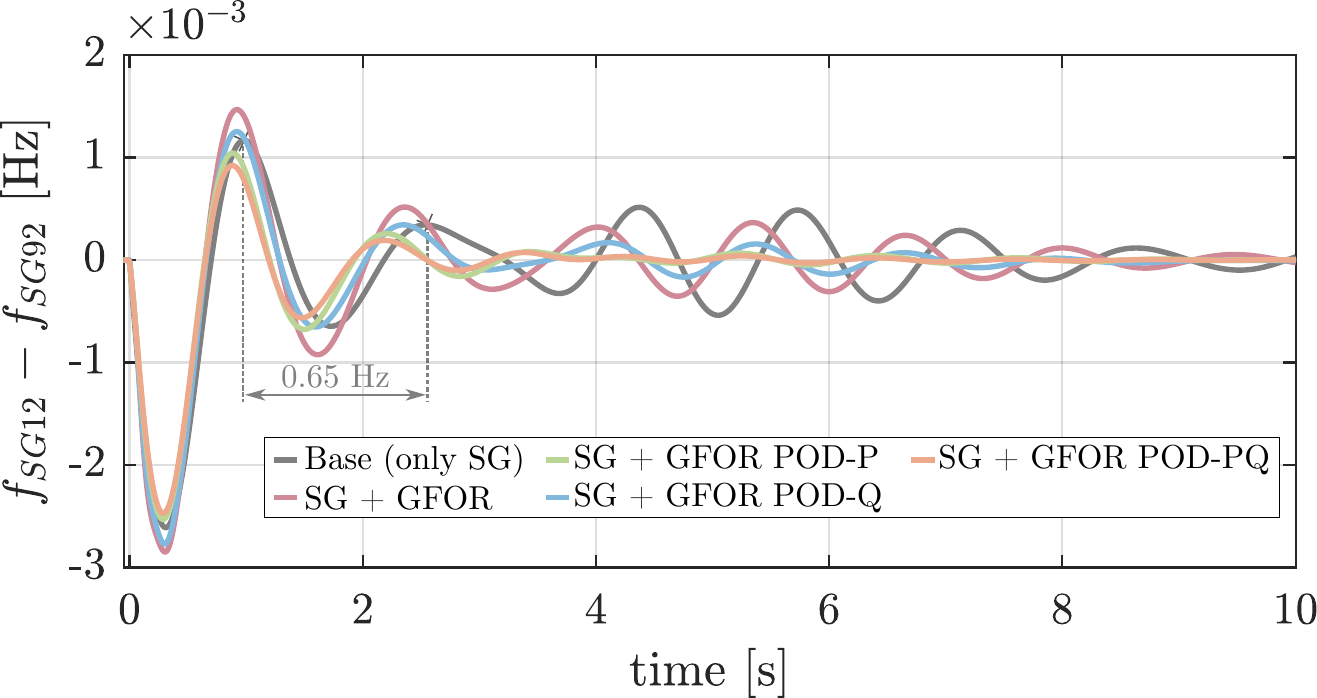}
\caption{Frequency difference between SG-12 and SG-92.}
\label{fig_118_w}
\end{figure}

First, the frequency difference between SGs in buses 12 and 92 is depicted in Fig.~\ref{fig_118_w}. It is also observed that the less-damped dynamic response corresponds to the Base case, while results are improved in the presence of GFORs (without POD controllers). POD controllers damp successfully the electromechanical modes, showing a dynamic response with a higher damping ratio. With POD-P controllers, the damping ratio of the electromechanical mode is higher than the one obtained with POD-Q controllers; while results are further improved using both together (POD-PQ). This is consistent with the small-signal analysis results from Fig.~\ref{fig_118_modes}. 

\begin{figure}[h!]
\centering
\includegraphics[width=\columnwidth]{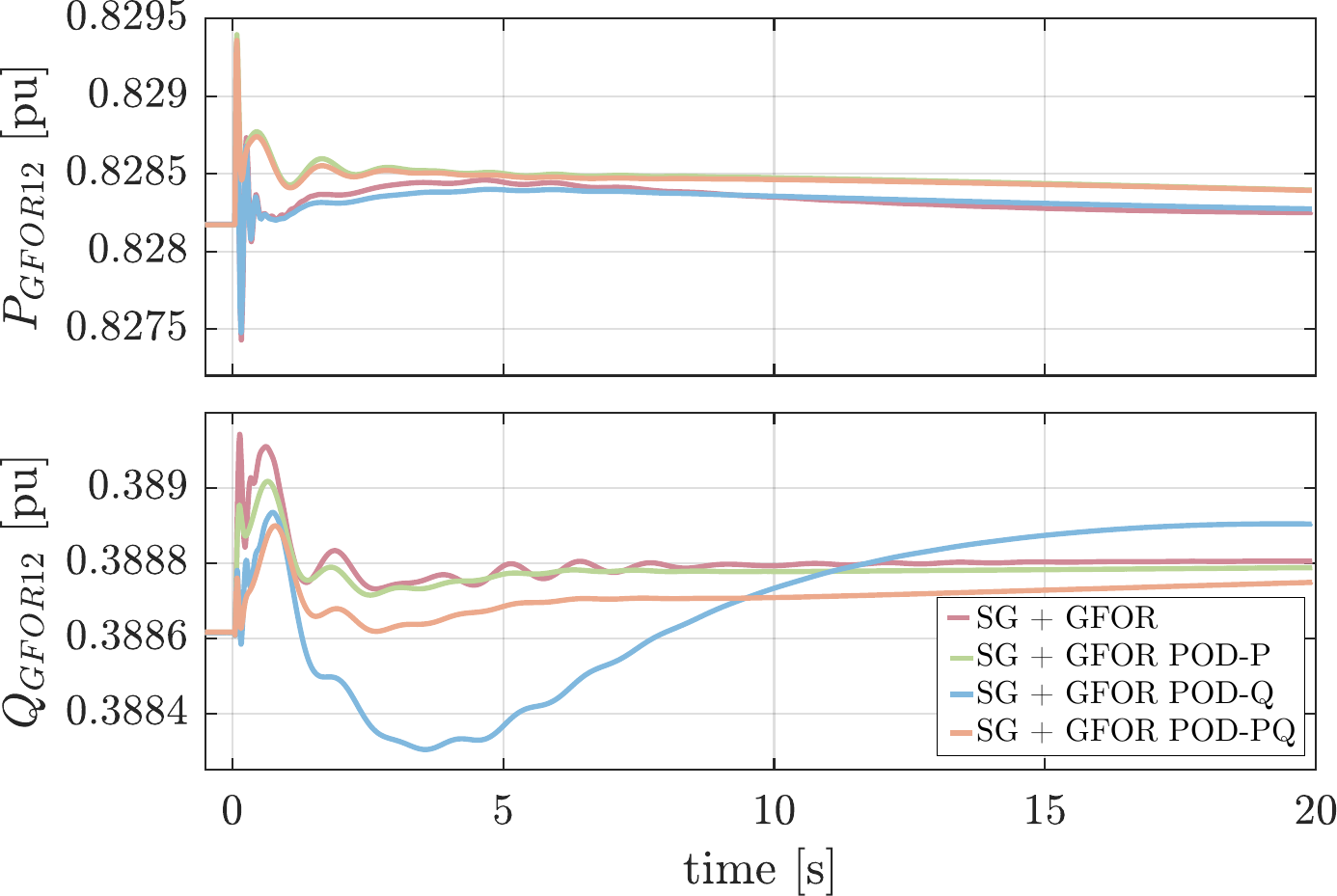}
\caption{Active and reactive power injection of GFOR in bus 12.}
\label{fig_118_PQ}
\centering
\includegraphics[width=\columnwidth]{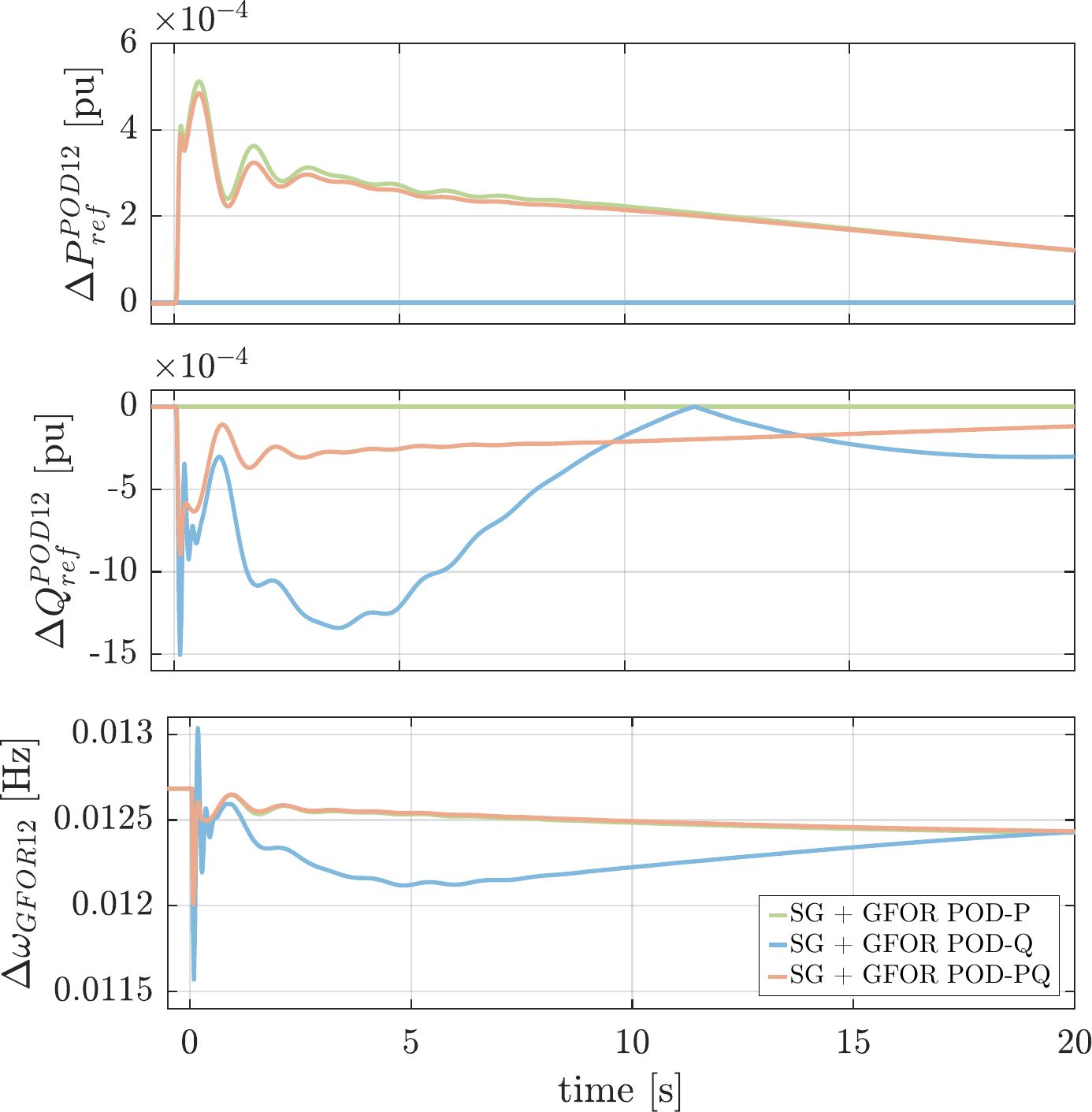}
\caption{Input (GFOR frequency deviation) and Output (Supplementary active and reactive power references) signals from POD control in bus 12.}
\label{fig_118_PQ_ref}
\end{figure}

Finally, Fig.~\ref{fig_118_PQ} shows the active and reactive power injection of GFOR in bus 12, and Fig.~\ref{fig_118_PQ_ref} shows the supplementary set-point values for POD controllers of GFOR in bus 12 (written in \unit{\pu} with respect to GFOR's nominal apparent power, which is \qty{90}{\mega\voltampere}) and the frequency deviations. After the event, P/Q injections of GFOR-12 change due to P-f/Q-V droop control and due to the effect of POD controllers. P/Q modulation due to POD controllers (Fig.~\ref{fig_118_PQ_ref}) contributes to damping of electromechanical oscillations in the system. \color{black}

\newpage
$ $ \\
$ $ \\
\section{Conclusion}
\noindent This paper has investigated the damping capability of \mbox{P-f/Q-V} droop-based Grid-Forming power converters (GFOR) and it proposed supplementary power oscillation damping controllers for the active- and reactive-power injections (POD-P and POD-Q, respectively) to damp electromechanical oscillations. Two case studies have been conducted, a two-area system (used for the design of the POD controllers and for fundamental analysis) and a large-scale system including converter-interfaced generation, providing a comparison of GFOR and POD performance through small-signal analysis and EMT time-domain simulations. The following conclusions were drawn: \color{black}
\begin{itemize}[leftmargin=*]
    \item This paper proposes POD-P and POD-Q controllers in \mbox{P-f/Q-V} droop-based GFOR converters using the frequency imposed by the GFOR as the input signal, which has a simple implementation and it eliminates the need for additional frequency measurements.
    \item This work shows that droop-based GFOR exhibits an inherent damping capability in the inter-area frequency region, although this effect depends intrinsically on the droop control parameters. Results could be significantly improved by incorporating the proposed POD-P and POD-Q controllers in GFOR converters.  
    \item Results show that POD-P controllers in GFORs could be more effective than POD-Q controllers, but the control actions of the former are directly linked to the primary energy source or energy storage of the power converter. On the other hand, POD-Q controllers in GFORs have the advantage that their control actions are not directly linked with the energy source of the power converters. It was also concluded that both POD-P and POD-Q controllers could be used simultaneously, achieving further improvements.  \color{black}
    \item Design methods using eigenvalue-sensitivity methods and synthetic test systems were applied to POD-P and POD-P controllers in GFOR power converters. The consistency of the observations from both case studies (a small test system and a large-scale power system) holds significant interest for design considerations. Particularly, that small synthetic systems could serve as valid benchmarks for the design of POD controllers in GFORs to be applied in large-scale power systems. 
\end{itemize}


\vspace{-1.5em}
\section*{Acknowledgment}
 The results presented are part of the R\&D project ref. K-ER53-IDGFC, which is a collaboration between CITCEA-UPC, Universidad Carlos III de Madrid and Red Eléctrica – Redeia and was funded by Red Eléctrica – Redeia. The authors thank Sergio Martinez Villanueva (Red Eléctrica – Redeia) for useful comments during the project.
 

\vspace{-1em}

\newpage
\appendix

\vspace{-1.5em}




\begin{table}[h]
\renewcommand{\arraystretch}{0.6}
\centering
\captionsetup{justification=centering, labelsep=newline, textfont=footnotesize}
\caption{Grid-forming converter parameters.}
\label{tab:gfor_param}
\begin{tabular}{ll}
\toprule
Rated values & \qty{230}{\kilo\volt} line-to-line, \qty{50}{\hertz} \\ 
Coupling filter & $R_c$ = \qty{0.005}{\pu}, $X_c$ = \qty{0.15}{\pu}, $B_c$ = \qty{0.15}{\pu} \\
                & $R_{cap}$ = $1/(3\cdot10\cdot \omega_b \cdot C_c)$  \\ \midrule
Current control & $\tau_{cc} = \qty{1}{\milli\second}$, $K_{P,cc} = L_c/\tau_{cc}$, $K_{I,cc} = R_c/\tau_{cc}$\\
Current saturation & $i_{q,max} = \qty{1.1}{\pu}$, $i_{d,max} = \qty{1.1}{\pu}$ \\ \midrule
AC voltage control & $\tau_{V_{ac}} = \qty{50}{\milli\second}$, $\xi = 0.707$, $\omega_{n}=4/(\tau_{V_{ac}} \cdot \xi)$ \\
                & $K_{P,V_{ac}} = 2 \cdot \xi \cdot \omega_{n} \cdot C_c \cdot 100$, $K_{I,V_{ac}} =  \omega_{n}^2 \cdot C_c$\\
Feed-forward filters & $\tau_{ff} = \qty{0.1}{\milli\second}$ \\ \midrule
P-f droop & $R_f = 0.05$, $\tau_P = \qty{0.1}{\second}$ \\
Q-V droop & $R_V = 0.067$, $\tau_Q = \qty{0.1}{\second}$\\ 
\bottomrule
\multicolumn{2}{l}{pu base: converter rated values}
\end{tabular}


\renewcommand{\arraystretch}{0.6}
\centering
\captionsetup{justification=centering, labelsep=newline, textfont=footnotesize}
\caption{Synchronous generator parameters for IEEE 118-bus system.}
\label{tab:sg_param}
\begin{tabular}{ll}
\toprule
Transformer & $R_{tr} = \qty{0.002}{\pu}$, $X_{tr} = \qty{0.1}{\pu}$ \\
Electrical machine & $R_s = \qty{0.0025}{\pu}$, $X_l = \qty{0.2}{\pu}$, $X_d = \qty{1.8}{\pu}$, \\
                   & $X'_d = \qty{0.3}{\pu}$, $X''_d = \qty{0.25}{\pu}$, $X_q = \qty{1.7}{\pu}$, \\
                   & $X'_q = \qty{0.55}{\pu}$, $X''_q = \qty{0.25}{\pu}$, $R_{snb} = \qty{300}{\pu}$ \\ 
                   & $T'_{q0} = \qty{0.4}{\second}$, $T''_{q0} = \qty{0.05}{\second}$, $T'_{d0} = \qty{8}{\second}$, $T''_{d0} = \qty{0.03}{\second}$\\ \midrule
Exciter AC4A & $K_A = 200$, $T_A = \qty{0.015}{\second}$, $T_B = \qty{10}{\second}$, $T_c = \qty{1}{\second}$ \\ \midrule
Governor-Turbine        & $R = 0.05$, $T_1 = T_2 = \qty{0}{\second}$, $T_3 = \qty{0.1}{\second}$, $K_1 = 0.3$ \\ 
IEEEG1                  & $K_2 = 0.4$, $K_3 = K_7 = K_4 = K_6 = K_8 = 0$ \\ 
                        & $K_5 = 0.3$, $T_4 = \qty{0.3}{\second}$, $T_5 = \qty{7}{\second}$, $T_6 = \qty{0.6}{\second}$, $T_7 = 0$\\\bottomrule
\multicolumn{2}{l}{pu base: SG rated values}
\end{tabular}
\end{table}

\FloatBarrier

\vspace{-1.7em}

\end{document}